\documentclass[secnumarabic,amssymb, nobibnotes, aps, prd]{revtex4}

\usepackage{amssymb}
\usepackage{amsmath}
\usepackage{bm}
\usepackage{graphicx,color}
\usepackage{color}

\def\be{\begin{equation}}
\def\ee{\end{equation}}

\def\te{\end{equation}}
\def\bea{\begin{eqnarray}}
\def\nn{\nonumber\\}
\def\tea{\end{eqnarray}}
\begin{document}

\title{Drag reduction by polymer additives from turbulent spectra}
\author{Esteban Calzetta}
\email[]{calzetta@df.uba.ar}
\affiliation{Departamento de Física, FCEN, Universidad de Buenos Aires, and IFIBA-CONICET, Pabellon I, 1428, Buenos Aires, Argentina}
\begin{abstract}
We extend the analysis of the friction factor for turbulent pipe flow reported by G. Gioia and P. Chakraborty (G. Gioia and P. Chakraborty, Phys. Rev. Lett. 96, 044502 (2006)) to the case where drag is reduced by polymer additives.
\end{abstract}

\maketitle

\section{Introduction}
When a fluid flows through a pipe of circular section and radius $R$ it experiences a pressure drop per unit length of pipe $dp/dx=-2\tau_0/R$, where $\tau_0$ is the stress at the wall. $\tau_0$ has units of energy density and is commonly parameterized as in the Darcy-Weisbach formula 

\be
\tau_0\equiv\frac f8\rho\:V^2
\label{Darcy}
\te
where $\rho$ is the density of the fluid and $V$ the mean velocity. The coefficient $f$ in eq. (\ref{Darcy}) is the so-called friction factor \cite{Sch50,MonYag71,Pop00,McZaSm05,YanJos08}.

For a given pipe, the friction factor is a function of Reynolds number $\mathrm{Re}$

\be
\mathrm{Re}=\frac{2RV}{\nu}
\label{Reynolds}
\te
where $\nu$ is the kinematic viscosity of the fluid (as distinct from the dynamic viscosity $\mu=\rho\nu$). It presents three power law-like regimes separated by transition regions. For laminar flows ($\mathrm{Re}<10^3$), $f=64/\mathrm{Re}$; for developed turbulent flows ($10^3<\mathrm{Re}<10^6$) it obeys the Blasius Law $f=0.3164/\mathrm{Re}^{1/4}$, and for larger values of Reynolds number it converges to an asymptotic value determined by the pipe roughness. 

G. Gioia and P. Chakraborty \cite{GioCha06} have presented a theoretical model whereby these features of the friction are easily derived from the Kolmogorov spectrum of homogeneous, isotropic turbulence. See also  \cite{GioBom02,Gol06,GiChBo06,MehPou08,GutGol08,Tran09,Gioi09,Cal09}. We shall refer to this as the momentum transfer model (MTM)

An immediate prediction of the MTM  is that, in situations where the turbulent spectrum deviates from the Kolmogorov form, the friction factor should also deviate from the Blasius Law. This prediction has been confirmed in the analysis of two dimensional flows \cite{GutGol08}. In this note, we wish to perform a similar analysis for three dimensional pipe flow in the presence of drag reducing polymer additives.

It is well known that adding a few parts per million of certain polymer additives to a fluid causes a drastic reduction in the friction factor \cite{Virk67,Lum69,Virk71,Virk71b,Ber78,Gen90,MCCO94,ThiBha96,Jov06,Oue07,PLB08,WhiMun08}. Use of this effect in improving the efficiency of oil and natural gas pipelines is widespread. There are indications that the phenomenon is not confined to pipe flow, but that the presence of the additives affect turbulence even in the homogeneous and isotropic limit \cite{PMP06}.

In this paper we will show that, given a turbulent spectrum for the pure solvent consistent with both the Blasius Law $f=.31/\mathrm{Re}^{1/4}$ and the Poiseuille Law $f=64/\mathrm{Re}$, then there is a deformation of this spectrum that reproduces the phenomenology of drag reduction, both in the asymptotic universal limits and with respect to the concentration dependence.

To incorporate the polymer to our model we shall adopt the theoretical framework provided by the so-called finitely extensive nonlinear elastic model supplemented by the Peterlin approximation (FENE-P model) \cite{Bir87,Ptas03}. To obtain the turbulent spectrum under homogeneous isotropic conditions, we shall map the nonlinear equations of the FENE-P model into an equivalent stochastic linear system, constructed to produce the right turbulent spectrum in the zero polymer concentration limit. By solving this linear problem, we shall find the spectrum as modified by finite polymer concentration, and we shall show that indeed it becomes universal in the large concentration, large Reynolds number limit. Moreover, by adopting the MTM, we will derive a power law dependence for the friction factor $f\approx \mathrm{Re}^{-1/2}$, close to the experimental result  \cite{Virk67}.

This paper is organized as follows: in next Section we review the MTM of the friction factor in the absence of the polymer. The original presentation of the MTM made contact with the Blasius and Strickler (for rough pipes) asymptotic regimes, but did not discuss in any detail the transition from the Blasius to the Poiseuille regimes \cite{GioCha06,Cal09}. To incorporate the Poiseuille regime within the MTM framework, we analyze the flow into a central region, where velocity fluctuations play an important role, and an outer region where fluctuations are negligible. We apply the MTM prescription to find the Reynolds stress at the boundary between these two regions, and then relate it to the stress at the wall by solving the Navier-Stokes equation in the outer region. We show that the resulting model gives the Poiseuille Law at low Reynolds numbers. At high Reynolds numbers the flow in the central region may be described as a Kolmogorov cascade, and in this case the model yields the Blasius Law.

In the following Section, we introduce the polymer. We show in the Appendix that the effect of the polymer in the outer region is negligible. To find the flow in the central region, we approximate the Navier-Stokes and FENE-P equations by a linear system driven by a stochastic force. The stochastic equation for the fluid  is chosen by requiring that, in the absence of the polymer, it reproduces the spectrum of fluctuations as described in refs. \cite{GioCha06,Cal09}. We then add the coupling to the polymer stress tensor as dictated by the FENE-P model. The equation for the polymer deformation tensor is a Hartree approximation to the original FENE-P equation.

The model leaves several parameters indetermined, the most important being the relaxation time of the polymer. We determine this parameter by requiring that at high Reynolds number the relaxation time for the polymer is proportional to the revolving time for the eddies that dominate momentum transfer in the MTM. These are the eddies whose size matches the width of the outer region. At low Reynolds numbers, the relaxation time regresses to its equilibrium value. Since the model is not sensitive to the details of the relaxation time dependence with respect to Reynolds number, we assume a simple interpolation formula that yields the proper asymptotic values. 

The result of this analysis is a friction factor - Reynolds number dependence containing five dimensionless parameters. We determine these parameters matching to experimental results, namely the Poiseuille and Blasius laws without the polymer, the Virk asymptote \cite{Virk67} at large polymer concentrations, and finally the detailed data presented in \cite{Virk67} for finite concentration. Having obtained a suitable set of parameters, we display the results in Section IV.

We conclude with a few brief remarks. In the appendix we discuss the FENE-P model in the outer region.

\section{The MTM applied to the pure solvent friction factor}
In the absence of the polymer, the dynamics of the solvent is described by the Navier-Stokes equation

\be
\left[\frac{\partial}{\partial t}-\nu_b\nabla^2\right]\mathbf{U}^p+\nabla_{q}\left[\mathbf{U}^p\mathbf{U}^q\right]+\nabla^pP=0
\label{eq4b}
\te
We are interested in a stationary flow within a straight pipe of circular section and radius $R$. Let $z$ be the coordinate along the pipe. The  flow may be decomposed into mean flow  and fluctuations as $\mathbf{U}^p=U\left(r\right)\hat{z}^p+\mathbf{u}^p$, where $\hat{z}^p$ is the unit vector in the $z$ direction. We call $V$ the average value of $U$ across the section of the pipe.  We define a dimensionless radial coordinate $\xi =r/R$.

We are interested in a high Reynolds number regime where the mean flow is very flat in the central region of the pipe, from $\xi=0$ to $\xi=\xi^*=1-\delta^*$, say, and there is an outer layer from there up to $\xi=1$. The precise form of the velocity profile in the central region is not a critical concern and we shall take it as simply flat,  with amplitude $U^*$. We adopt the convention of computing the Reynolds number as if the central flow filled the whole pipe; this is only a matter of convenience. The central flow  is characterized by a Reynolds number

\be
\mathrm{Re}^*=\frac{2U^*R}{\nu}
\label{1}
\te
In the outer layer we neglect the fluctuating velocity $\mathbf{u}^p$. Then the Navier-Stokes equations reduce to

\be
\frac{d}{d\xi}\frac1{\xi}\frac{d}{d\xi}\xi\frac{dU}{d\xi}=0
\label{2}
\te
The solution that vanishes at $\xi=1$ reads

\be
U\left[\xi\right]=U^*\left[a\left(1-\xi^2\right)+b\ln\left[\xi\right]\right]
\label{3}
\te
Asking the mean velocity profile to be continuous we get

\be
1=a\left(1-\xi^{*2}\right)+b\ln\left[\xi^*\right]
\label{4}
\te
We also have the shear stress at $\xi^*$, namely 

\be
\tau_{turb}\equiv\frac{\rho\nu U^*}R\chi^*
\label{5}
\te
where

\be
\chi^*=-\frac1{U^*}\frac{dU}{d\xi}\left[\xi^*\right]=2a\xi^*-\frac b{\xi^*}
\label{6}
\te
We can write the constants $a$ and $b$ in terms of $\chi^*$ as

\bea
a&=&\frac1{\Delta\left[\xi^*\right]}\left[1+\xi^*\ln\left[\xi^*\right]\chi^*\right]\nn
b&=&\frac{\xi^*}{\Delta\left[\xi^*\right]}\left[2\xi^*-\left(1-\xi^{*2}\right)\chi^*\right]\nn
\Delta\left[\xi^*\right]&=&1-\xi^{*2}+2\xi^{*2}\ln\left[\xi^*\right]
\label{7}
\tea
Our interest is to find the average velocity, which enters in the Reynolds number eq. (\ref{Reynolds}) 

\be
V=\xi^{*2}U^*+\int_{\xi^{*2}}^1dx\;U\left[\sqrt{x}\right]=U^*h\left[\delta^{*}\right]
\label{11a}
\te

\be
h\left[\delta^{*}\right]=\frac {\left[1-\xi^{*2}\right]^2}{2\Delta\left[\xi^*\right]}\left\{1+\chi^* \xi^*\left[1+\left(1+\xi^{*2}\right)\frac{\ln\left[\xi^*\right]}{\left(1-\xi^{*2}\right)}\right]\right\}
\label{11}
\te
and the stress at the wall

\be
\tau_{0}\equiv\frac{\rho\nu U^*}R\chi
\label{13}
\te
where

\bea\chi&=&-\frac1{U^*}\frac{dU}{d\xi}\left[1\right]=2a-b
\label{14}
\tea
If we parameterize $\tau_0$ as in eq. (\ref{Darcy}), then the friction factor

\be
f=\frac{8\tau_0}{\rho V^2}=\frac{64}{\mathrm{Re}}\frac{\chi U^*}{4V}
\label{15}
\te
If $\xi^*\to 0$ then $a\to 1$, $b\to 0$, $V\to U^*/2$ and $\chi\to 2$, so we recover the Poiseuille Law. 

In the general case, we need to relate $U^*$ and $\chi^*$ to $\xi^*$ to obtain the friction factor - Reynolds number dependence in parametric form. These relations are provided by the momentum transfer model (MTM). In the central region, each scale $\delta$ is associated to a velocity $u\left[\delta, U^*\right]$. The MTM claims that

a) the width $\delta^*$ of the outer layer is also the Kolmogorov scale of the central region, namely the scale at which the Reynolds number is $1$

\be
\frac{R\delta^*u\left[\delta^*, U^*\right]}{\nu}=1
\label{9}
\te
Note that in the original presentation of the MTM only a proportionality between $\delta^*$ and the Kolmogorov scale is required \cite{GioCha06}. We have adopted the more restrictive criterion (\ref{9}) to simplify the discussion below.

Let us write

\be
u\left[\delta^*, U^*\right]=U^*v\left[\delta^*\right]
\label{210}
\te
then eq. (\ref{9}) may be rewritten as

\be
\mathrm{Re}^*=\frac 2{\delta^*v\left[\delta^*\right]}
\label{211}
\te

b) the shear stress at the boundary of the central region is

\be
\tau_{turb}=\rho U^* u\left[\delta^*,U^*\right]
\label{8}
\te
As in case (a), we have opted for postulating an equality where the original MTM only asks for proportionality \cite{GioCha06}. Therefore

\be
\chi^*=\frac{1}{\delta^*}
\label{10}
\te
and the constants read

\bea
a&=&\frac{1}{\Delta\left[\xi^*\right]}\left[1+\frac{\xi^*}{\delta^*}\ln\left[\xi^*\right]\right]\nn
b&=&\frac{-\xi^*}{\Delta\left[\xi^*\right]}\left[1-\xi^*\right]
\label{7b}
\tea
leading to

\be
\chi=\frac1{\delta^*\xi^*}\left\{1-\frac{\delta^{*3}\left(1+\xi^*\right)}{\Delta\left[\xi^*\right]}\right\}\equiv\frac{\chi_0\left[\delta^*\right]}{\delta^*}
\label{16}
\te
We may now write the friction factor - Reynolds number dependence in parametric form

\bea
\mathrm{Re}&=&\zeta\left[\delta^*\right]\nn
f&=&\varphi\left[\delta^*\right]
\label{parametric}
\tea
where

\bea
\zeta\left[\delta^*\right]&=&\frac{2h\left[\delta^*\right]}{\delta^*v\left[\delta^*\right]}\nn
\varphi\left[\delta^*\right]&=&\frac{8\chi_0\left[\delta^*\right]}{h\left[\delta^*\right]^2}v\left[\delta^*\right]
\label{parametric2}
\tea
In this formulae we already have explicit expressions for $h\left[\delta^*\right]$ (cfr. eq. (\ref{11})) and $\chi_0\left[\delta^*\right]$ (cfr. eq. (\ref{16})), but we need a detailed model of the velocity fluctuations in the central region to derive $v\left[\delta^*\right]$. This shall be our concern in the rest of the paper.

We have already remarked, however, that if $\delta^*\to 1$ then the parametric equations (\ref{parametric}) reproduce the Poiseuille law, and it can be seen by inspection that a Kolmogorov scaling $v\left[\delta^*\right]\propto\delta^{*1/3}$ when $\delta^*\to 0$ will produce the Blasius law, if appropriate values for the several constants in the theory may be found. Therefore we may be confident that our model successfully reproduces the limiting behaviors.

\section{Solvent - polymer interaction}
\subsection{The FENE-P model}
In this section we consider the modifications of the above picture due to the addition of the polymer. We shall adopt the so-called FENE-P model. The fluid velocity  obeys the incompressibility condition $\nabla_{p}\mathbf{U}^p=0$ and a modified Navier-Stokes equation

\be
\left[\frac{\partial}{\partial t}-\nu\nabla^2\right]\mathbf{U}^p+\nabla_{q}\left[\mathbf{U}^p\mathbf{U}^q\right]+\nabla^pP=\frac1{\rho}\nabla_{q}\left(\rho_pT^{pq}\right)
\label{eq4}
\te
where $P$ is the pressure, $\rho$ the fluid density, $\rho_p$ the polymer density and $T^{pq}$ a polymer stress. $T^{pq}$ is modeled in terms of the polymer deformation tensor $\Lambda^{pq}$ as

\be
T^{pq}=\omega_{free}^2h\left(\Lambda\right)\Lambda^{pq}
\label{tpq}
\te
where $\omega_{free}$ is the frequency of free oscillations of the molecule, $\Lambda=(1/3)\mathrm{tr}\Lambda^{pq}$ and $h$ is some function that is close to one under equilibrium conditions $\Lambda=\Lambda_{eq}$ and diverges as $\Lambda$ approaches maximum elongation $\Lambda=\Lambda_{max}$.

The evolution of the deformation tensor is determined by the drag from the fluid and the polymer elasticity. Neglecting the inertia of the molecule, we get

\be
\left[\frac{\partial}{\partial t}+\mathbf{U}^r\nabla_{r}\right] \Lambda^{pq}=\left(\nabla_{r}\mathbf{U}^p\right)\Lambda^{rq}+\Lambda^{pr}\left(\nabla_{r}\mathbf{U}^q\right)-t_{S}  \left(T^{pq}-T^{pq}_{eq}\right)
\label{FENEP}
\te
$t_{S}$ is the time scale in which a freely moving bead from the polymer would come to rest with respect to the fluid.

\subsection{Flow in the central region}
We shall make the approximation that the flow in the outer region is not affected by the polymer. This issue is further discussed in the Appendix. Under this approximation, the analysis in the previous section remains valid, and the only effect of the polymer is changing the functional form of $u\left[\delta, U^*\right]$ in eqs. (\ref{9}) and (\ref{8}). To find this, we only need to consider the fluctuating part of the velocity. We shall consider only the homogeneous, isotropic case, since the behavior of the fluid in this case determines the friction factor in the MTM. To take advantage of the symmetries of the problem, we shall decompose the deformation tensor into its scalar and traceless parts

\be
\Lambda^{pq}=\Lambda \delta^{pq}+\lambda^{pq}
\label{scalarplustraceless}
\te
$\lambda^{p}_p=0$. Moreover, we shall assume that $\Lambda $ is both space and time independent. The stress tensor is decomposed into a similar way

\be
T^{pq}=T \delta^{pq}+\omega_{free}^2h\left(\Lambda\right)\lambda^{pq}
\label{scalarplustraceless2}
\te
where 

\be
T=\omega_{free}^2h\left(\Lambda\right)\Lambda
\label{T}
\te
Taking the trace of eq. (\ref{FENEP}) we obtain 

\be
3t_{S}  \left(T-T_{eq}\right)=2\lambda^{pq}\nabla_{p}\mathbf{u}_q
\label{Hartree}
\te
Subtracting the trace from eq. (\ref{FENEP}) we get

\be
\frac{\partial}{\partial t} \lambda^{pq}=\Lambda\left(\nabla^{p}\mathbf{u}^q+\nabla^{q}\mathbf{u}^p\right)-\frac1{\tau}\lambda^{pq}+S^{pq} 
\label{tracelessFENEP}
\te
where

\be
\frac1{\tau}=\omega_{free}^2h\left(\Lambda\right)t_{S}=\frac{Tt_{S}}{\Lambda} 
\label{tau}
\te
and

\be
S^{pq}=\left(\nabla_{r}\mathbf{u}^p\right)\lambda^{rq}+\lambda^{pr}\left(\nabla_{r}\mathbf{u}^q\right)-\frac23\lambda^{rs}\left(\nabla_{r}\mathbf{u}_s\right)\delta^{pq}-\mathbf{u}^r\nabla_{r}\lambda^{pq}
\label{spq}
\te
whose ensemble average must be zero from the symmetries of the problem. We shall neglect $S^{pq}$ in what follows.

\subsection{Equivalent linear stochastic model}

It is clear that the full Navier-Stokes is too complex for analysis, unless numerically \cite{Ptas03}. To make progress, we shall substitute the Navier-Stokes equation by a linear stochastic one, devised to give the right spectrum in the absence of the polymer. 

Let us begin by Fourier decomposing the fluid velocity

\be
\mathbf{u}^p\left(\mathbf{x},t\right)=\int\:\frac{d\mathbf{k}}{\left(2\pi\right)^3}\:\frac{d\omega}{\left(2\pi\right)}\:e^{i\left[\mathbf{k}\mathbf{x}-\omega t\right]}\mathbf{u}^p_{\mathbf{k}}\left[\omega\right]
\label{Fourier}
\te
We postulate for the Fourier components a dynamic equation

\be
\left[-i\omega+\sigma_k\right]\mathbf{u}^p_{\mathbf{k}}=F^p_{\mathbf{k}}\left[\omega\right]
\label{linear}
\te
where $F^p_{\mathbf{k}}$ is a Gaussian random source with self correlation

\be
\left\langle F^p_{\mathbf{k}}\left[\omega\right]F^q_{\mathbf{k'}}\left[\omega'\right]\right\rangle=\left(2\pi\right)^6\delta\left(\mathbf{k}+\mathbf{k'}\right)\delta\left(\omega+\omega'\right)\Delta^{pq}_{\mathbf{k}}{N}_k
\label{self}
\te

\be
\Delta^{pq}_{\mathbf{k}}=\delta^{pq}-\frac{\mathbf{k}^p\mathbf{k}^q}{k^2}
\label{delta}
\te
A representation like this may be derived from the functional approach to turbulence, where the left hand side of eq. (\ref{linear}) is identified as the inverse retarded propagator, and the self-correlation eq. (\ref{self}) is given by a self-energy\cite{MCCO94,Cal09b}. We shall be content to propose simple expressions for $\sigma_k$ and $\mathbf{N}_k$ to reproduce the known turbulent spectrum.

In the inertial range, we expect $\sigma_k$ and $\mathbf{N}_k$ to depend on the only dimensionful parameter $\epsilon$, which is  the energy flux feeding the Richardson cascade. On dimensional grounds \cite{YII02}

\be
\sigma_k=\nu_0\left(k^2\epsilon\right)^{1/3}
\label{sigmainertial}
\te
where $\nu_0$ is a dimensionless constant to be determined presently. The turbulent spectrum $E_0\left[k\right]$ (where the $0$ subscript denotes that this is the spectrum in the absence of the polymer) is defined from the mode decomposition of the turbulent energy

\be
\left\langle \mathbf{u}^2\left[\mathbf{x},t\right]\right\rangle=2\int_{0}^{\infty}\:dk\:E_0\left[k\right]
\label{global2}
\te
Explicitly

\be
E_0\left[k\right]=k^2\frac{N_k}{\sigma_k}
\label{nsobresigma}
\te
so we recover the Kolmogorov spectrum $E_0\left[k\right]=C_K\epsilon^{2/3}k^{-5/3}$, where $C_K\approx 1.5$ is the so-called Kolmogorov constant \cite{MCCO94},  provided

\be
N_k=\nu_0C_K\frac{\epsilon}{k^3}
\label{noise}
\te
The ansatz eq. (\ref{sigmainertial}) for $\sigma_k$ is equivalent to a scale dependent viscosity $\nu_k=k^{-2}\sigma_k=\nu_0\epsilon^{1/3}k^{-4/3}$. Under Kolmogorov scaling the velocity associated to a scale $\delta=k^{-1}$ is $u_k=\sqrt{3C_K}\left(\epsilon k^{-1}\right)^{1/3}$. We find the Reynolds number of the effective linear theory as $\mathrm{Re}_{eff}=u_k/k\nu_k=\sqrt{3C_K}/\nu_0$. Identifying this with the physical Reynolds number of the central region we get 

\be
\nu_0=\frac{\sqrt{3C_K}}{\mathrm{Re}^*}
\label{nucero}
\te
This simple picture must be modified to account for the dissipative range. In the dissipative range  fluctuations are strongly suppressed

\be
N_k=\frac{\sqrt{3C^3_K}}{\mathrm{Re}^*}\frac{\epsilon}{k^3}e^{-\beta \delta^*R k}
\label{sigmadissipative}
\te
with $\beta\approx 1/2$ a dimensionless number \cite{Cal09}. The strong suppression of fluctuations dispenses with further discussion of $\sigma_k$ in this range; on general grounds we expect it will approach its bare value $\nu k^2$, but for simplicity we shall use the inertial form eq. (\ref{sigmainertial}) in the calculations below. 

At very long wavelengths the spectrum must turn over and approach the von Karman spectrum $E\left[k\right]\propto k^4$ \cite{Kar48}. To obtain this we add one further factor, turning the noise correlation into

\be
N_k=\frac{\sqrt{3C^3_K}}{\mathrm{Re}^*}\frac{\epsilon R^{17/3}k^{8/3}}{\left[\gamma +\left(Rk\right)^2\right]^{17/6}}e^{-\beta \delta^*R k}
\label{karmanmod}
\te
This yields the same spectrum as assumed in ref. \cite{GioCha06}.

\subsection{Solving the effective linear model}
Adopting the linear effective model as  a suitable description of the eddy dynamics, and for a constant polymer density $\rho_p=c\rho$, we get, instead of eqs. (\ref{eq4}) and (\ref{FENEP}), the linear system

\be
\left[-i\omega+\sigma_k\right]\mathbf{u}^p_{\mathbf{k}}-\frac {ic}{\tau t_{S}}\mathbf{k}_q\lambda_{\mathbf{k}}^{pq}=F^p_{\mathbf{k}}\left[\omega\right]
\label{linear2}
\te

\be
\left[-i\omega+\frac1{\tau}\right] \lambda^{pq}_{\mathbf{k}}=i\Lambda\left(\mathbf{k}^{p}\mathbf{u}^q_{\mathbf{k}}+\mathbf{k}^{q}\mathbf{u}^p_{\mathbf{k}}\right)
\label{linearFENEP}
\te
Eliminating $\lambda^{pq}_{\mathbf{k}}$ we obtain a second order equation for $\mathbf{u}^p_{\mathbf{k}}$ (we also use the incompressibility constraint $\mathbf{k}_{p}\mathbf{u}^p_{\mathbf{k}}=0$)

\be
P\left[\omega\right]\mathbf{u}^p_{\mathbf{k}}=-\left[-i\omega+\frac1{\tau}\right]F^p_{\mathbf{k}}\left[\omega\right]
\label{secondorder}
\te

\bea
P\left[\omega\right]&=&\omega^2+i\omega\left[\sigma_k+\frac1{\tau}\right]-\frac1{\tau}\left[\sigma_k+\frac {c}{ t_{S}}\Lambda k^2\right]\nn
&=&\left(\omega -s_+\right)\left(\omega -s_-\right)
\label{secular}
\tea
$s_{\pm}$ are the free frequencies of the system and are given by

\be
s_{\pm}=\frac12\left[-i\left(\sigma_k+\frac1{\tau}\right)\pm\sqrt{\frac {4c\Lambda k^2}{\tau t_{S}}-\left(\sigma_k-\frac1{\tau}\right)^2}\right]
\label{free}
\te
We see that there are two different flow regimes. When the discriminant in eq. (\ref{free}) is negative, both eigenfrequencies are pure imaginary. The imaginary part of both is always negative, so the flow is always stable. We shall call this the overdamped regime. This regime prevails in the energy range (where $k\to 0$, and therefore also $\sigma_k$) and in the dissipative range, where $\sigma_k$ is very large.

On the other hand, precisely because $\sigma_k$ goes from $0$ at $k=0$ to a very large value in the dissipative range, there must be some interval where $\sigma_k\approx 1/\tau$ and the discriminant is positive. In this regime the free frequencies have nonzero real parts, although they still describe damped oscillations. We shall call this the underdamped regime. As the concentration $c$ grows, the underdamped range expands and essentially becomes identical with the inertial range.

From the solution to eq. (\ref{secondorder}) and the noise self-correlation eq. (\ref{self}) we identify the spectrum in the presence of the polymer as

\be
E\left[k\right]=k^2N_k\left(\frac1{\tau^2}J\left[k\right]+J_2\left[k\right]\right)
\label{spectrum}
\te
where

\be
J\left[k\right]=\int\:\frac{d\omega}{\pi}\:\frac1{P\left[\omega\right]P\left[-\omega\right]}
\label{jota}
\te
and

\be
J_2\left[k\right]=\int\:\frac{d\omega}{\pi}\:\frac{\omega^2}{P\left[\omega\right]P\left[-\omega\right]}
\label{jotados}
\te
Actually these integrals are related

\be
J_2\left[k\right]=\left[\sigma_k+\frac {c\Lambda k^2}{ t_{S}}\right]\frac {J\left[k\right]}{\tau}
\label{relation}
\te
Evaluating the integral

\be
J\left[k\right]=\frac{i}{s_+s_-\left(s_++s_-\right)}=\frac{\tau^2}{\sigma_k}\frac1{\left[1+\sigma_k\tau\right]\left[1+\frac {c\Lambda k^2}{ t_{S}\sigma_k}\right]}
\label{integral2}
\te
and so the spectrum is

\be
E\left[k\right]=\frac{k^2N_k}{\sigma_k}\frac{\left[1+\sigma_k\tau+\frac {c\tau\Lambda k^2}{ t_{S}}\right]} {\left[1+\sigma_k\tau\right]\left[1+\frac {c\Lambda k^2}{ t_{S}\sigma_k}\right]}
\label{spectrum2}
\te

\subsection{Identifying the free parameters}
To give meaning to eq. (\ref{spectrum2}) we must know the way parameters such as $\epsilon$, $\tau$ and $\Lambda$ depend on $\delta^*$. The determination of these parameters is the subject of this section.

Let us begin with the expression of $u\left[\delta^*,U^*\right]$ in terms of the spectrum (cfr. eq. (\ref{global2}))

\be
u\left[\delta^*,U^*\right]^2=2\int_{\left(R\delta^*\right)^{-1}}^{\infty}\:dk\:E\left[k\right]
\label{300}
\te
writing $x=\delta^*Rk$ this becomes

\be
u\left[\delta^*,U^*\right]^2=2C_K\delta^{*2/3}\left(R\epsilon\right)^{2/3}\int_{1}^{\infty}\:dx\:\frac{x^4e^{-x/2}}{\left[\gamma\delta^{*2}+x^2\right]^{17/6}}\frac{\left[1+\sigma_k\tau+\frac {c\tau\Lambda k^2}{ t_{S}}\right]} {\left[1+\sigma_k\tau\right]\left[1+\frac {c\Lambda k^2}{ t_{S}\sigma_k}\right]}
\label{301}
\te
To obtain $\epsilon$, we observe that energy is fed into the Richardson cascade at a scale $R\xi^*$. Therefore we expect

\be
\epsilon=\left(\frac{\kappa}{\sqrt{3C_K}}\right)^3\frac{U^{*3}}{\xi^*R}
\label{302}
\te
with $\kappa$ a dimensionless parameter to be determined. This leads to 

\be
\sigma_k=\frac{\kappa}2\frac{\nu}{R^2}\frac{x^{2/3}}{\delta^{*2/3}\xi^{*1/3}}
\label{303}
\te
To find $\tau$,  we expect that at large Reynolds numbers $\tau$ will be  proportional to the revolving time for eddies of size $\delta^*$, namely 

\be
\tau=\frac{\alpha' R\delta^*}{ u\left[\delta^*,U^*\right]}\equiv\frac{ \alpha'  R^2\delta^{*2}}{\nu}
\label{100c}
\te
where $\alpha'$ is a new free parameter. For lower Reynolds numbers, we expect $\tau$ will regress to its equilibrium value $1/\omega_{free}^2t_S$. To interpolate between these regimes, we assume

\be
\frac1{\tau}=\omega_{free}^2t_S\left(1+\frac{\delta_0^2}{ \delta^{*2}}\right)
\label{304}
\te
where

\be
\delta_0^2=\frac{\nu}{ \alpha'  R^2\omega_{free}^2t_S}
\label{304b}
\te
$\Lambda$ is related to $\tau$ through eq. (\ref{tau}). If we neglect the ratio $\Lambda_{eq}/\Lambda_{max}$ and parameterize 

\be
h\left(\Lambda\right)\approx\frac1{1-\frac{\Lambda}{\Lambda_{max}}}
\label{305}
\te
then

\be
\Lambda =\frac{\Lambda_{max}}{1+\frac{ \delta^{*2}}{\delta_0^2}}
\label{306}
\te
This leads to

\be
\sigma_k\tau=\alpha\frac{\delta^{*4/3}}{\xi^{*1/3}}\frac{\delta_0^2}{\delta_0^2+\delta^{*2}}x^{2/3}
\label{307}
\te

\be
\frac {c\tau\Lambda k^2}{ t_{S}}=\frac{c}{c_0}\frac{\delta_0^4}{\left[\delta_0^2+\delta^{*2}\right]^2}x^{2}
\label{308}
\te

\be
\frac {c\Lambda k^2}{ t_{S}\sigma_k}=\frac1{\alpha}\frac{c}{c_0}\frac{\xi^{*1/3}}{\delta^{*4/3}}\frac{\delta_0^2}{\left[\delta_0^2+\delta^{*2}\right]}x^{4/3}
\label{309}
\te
where

\be
\alpha =\frac{\kappa\alpha'}2
\label{309b}
\te

\be
c_0=\frac{\nu t_S}{\alpha\Lambda_{max}}
\label{310}
\te
From eqs. (\ref{301}) and (\ref{210}) we get

\be
v\left[\delta^*\right]=\kappa\frac{\delta^{*1/3}}{\xi^{*1/3}}\left\{\frac 23\int_{1}^{\infty}\:dx\:\frac{x^4e^{-x/2}}{\left[\gamma\delta^{*2}+x^2\right]^{17/6}}\frac{\left[1+\frac{\delta^{*4/3}}{\xi^{*1/3}}\frac{\alpha\delta_0^2}{\delta_0^2+\delta^{*2}}x^{2/3}+\frac{c}{c_0}\frac{\delta_0^4}{\left[\delta_0^2+\delta^{*2}\right]^2}x^{2}\right]} {\left[1+\frac{\delta^{*4/3}}{\xi^{*1/3}}\frac{\alpha\delta_0^2}{\delta_0^2+\delta^{*2}}x^{2/3}\right]\left[1+\frac1{\alpha}\frac{c}{c_0}\frac{\xi^{*1/3}}{\delta^{*4/3}}\frac{\delta_0^2}{\left[\delta_0^2+\delta^{*2}\right]}x^{4/3}\right]}\right\}^{1/2}
\label{400}
\te
We see  that the solution depends on the parameters $\kappa$, $\gamma$, $\alpha$, $c_0$ and $\delta_0$. In principle, each of these could be a function of Reynolds number or other dimensionless combinations, This would turn the parametric relations eq. (\ref{parametric}) into implicit equations. For simplicity, we shall model them as  constants.

Let us check that our model shields the appropriate limiting behavior. To begin with, to obtain $\mathrm{Re}\to 0$ when $\delta^*\to 1$ we need that $v\left[\delta^*\right]$ should diverge in this limit, which indeed it is a result of eq. (\ref{400}). This does not contradict the fact that $U^*$ remains finite because in this limit the scale $R\delta^*$ is larger than the scale $R\xi^*$ at which energy is injected into the fluctuations. In this limit, of course, the fluctuations themselves cannot be regarded as turbulent in any conventional sense, and our model should be regarded as an extrapolation which ``saves the appearances''.

If $c=0$ and $\delta^*\to 0$ then eq. (\ref{400}) predicts $v\left[\delta^*\right]\propto\delta^{*1/3}$ and leads to the Blasius Law. We see that $\gamma\not= 0$ is necessary to obtain the ``hump'' feature in the friction factor plot \cite{GioCha06,Cal09}.

If $\delta^*\to 0$ but $c$ is large, then eq. (\ref{400}) predicts $v\left[\delta^*\right]\propto\delta^{*}$ and then the parametric relations become $f\propto \mathrm{Re}^{-1/2}$, in reasonable agreement with Virk's asymptote \cite{Virk67,Trinh10,RoyLar05,BraRob05,YanDou08}. We see however that even for large concentrations drag reduction will be very small as long as $\delta^*\gg\delta_0$ 

\section{Results}
In this section, we shall use the previous analysis to obtain concrete estimates of the friction factor. We adopt the following values for the free parameters: $\kappa=0.02$, $\gamma=40$, $\alpha=0.05$, $1/\delta_0^2=5000$ ($\delta_0=0.014$) and $c_0=43.6/0.00035=124600$ wppm

\begin{figure}[htp]
\centering
\includegraphics[scale=.8]{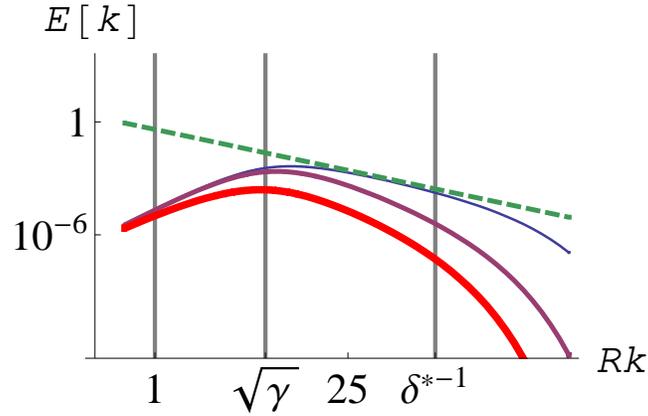}
\caption[k1] {[Color online] The concentration dependent spectra for $\mathrm{Re}=10^5$. The full lines, from the top down, correspond to $c/c_0=0$, $0.01$ and $1$; the concentration scale $c_0=124600$ wppm is defined in the text. The dashed line represents the Kolmogorov spectrum. We also show the beginning and the end of the inertial range in the pure fluid limit}
\label{k1}
\end{figure}

In fig. (\ref{k1}) we show the concentration dependent spectra for $\mathrm{Re}=10^5$ and $c/c_0=0$, $0.01$ and $1$. 

\begin{figure}[htp]
\centering
\includegraphics[scale=.8]{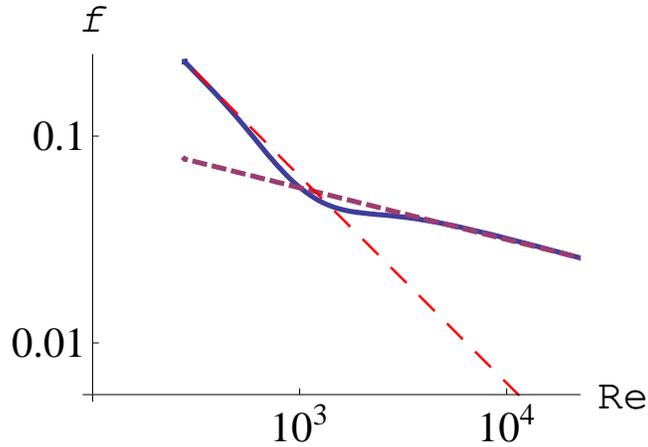}
\caption[k2] {[Color online] The friction factor  in the absence of the polymer (full line). The straight dashed lines represent the Blasius (short dashes) and Poiseuille Laws (long dashes)}
\label{k2}
\end{figure}

We now discuss the solution to the parametric equations (\ref{parametric}). In fig. (\ref{k2}) we show the friction factor for the pure fluid. The transition from the Blasius to the Poiseuille regimes is clearly seen. To obtain a more accurate fit to experimental data would require the introduction of a more complex spectrum and is not relevant to the discussion of drag reduction.

\begin{figure}[htp]
\centering
\includegraphics[scale=.8]{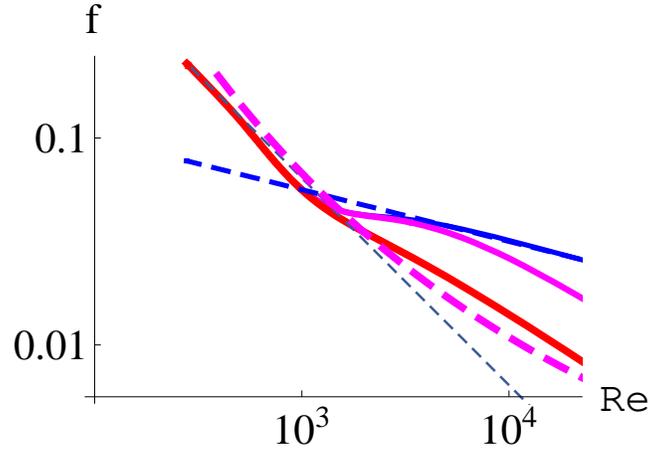}
\caption[k3] {[Color online] The friction factor for non-zero concentration. The full lines, from the top down, correspond to $c/c_0=0$, $0.01$ and $1$. The dashed lines, from the top down, correspond to the Blasius, Virk and Poiseuille Laws.}
\label{k3}
\end{figure}

In fig. (\ref{k3}) we add, to the $c/c_0=0$ line in fig. (\ref{k2}), the friction factor dependence for $c/c_0=0.01$ and $1$. We also add the Virk asymptote for comparison.

It is convenient to introduce the Prandtl-von Karman variables $X=\ln\left[\mathrm{Re}\sqrt{f}\right]$ and $Y=1/\sqrt{f}$. In these coordinates the Poiseuille Law becomes $Y=e^X/64$. In the turbulent regime, $Y$ is described by the Prandtl Law $Y=0.81X-0.8$. While the Prandtl and Blasius Laws, which reads $Y=\left(.31\right)^{4/7}e^{X/7}$, have very different mathematical expression, they give equivalent  results in the range of Reynolds numbers we are considering.

\begin{figure}[htp]
\centering
\includegraphics[scale=.8]{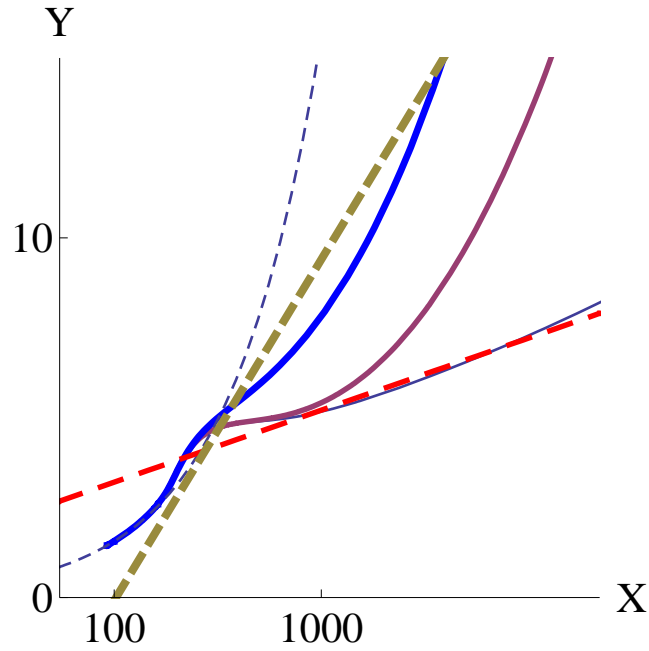}
\caption[k4] {[Color online] The friction factor from Fig. eq. (\ref{k3}) in Prandtl-von Karman coordinates. The dashed lines are, from the bottom up, the Prandtl, Virk and Poiseuille Laws}
\label{k4}
\end{figure}

In fig. (\ref{k4}) we reproduce fig.(\ref{k3}) in Prandtl-von Karman coordinates. We have also added the Virk asymptote $Y=4.12X-19.06$ \cite{Virk67,Trinh10,RoyLar05}. 

\begin{figure}[htp]
\centering
\includegraphics[scale=.8]{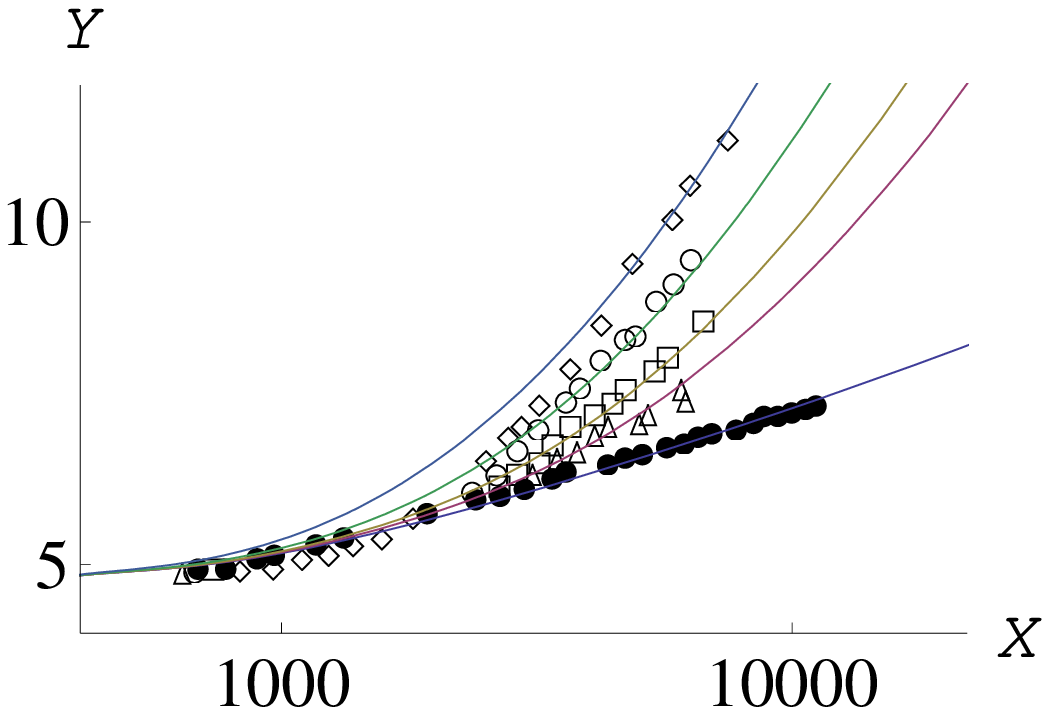}
\caption[k5] {[Color online] The friction factor for small polymer concentration. The symbols represent the data presented in Fig. 2a of ref. \cite{Virk71}. They correspond to measurements of the friction factor in solutions of a single polymer (N750) in five different concentrations: pure solvent (full circles), and $43.6$ (triangles), $98.6$ (squares), $296$ (circles) and $939$ (diamonds) w.p.p.m..For the theoretical lines we have used the parameters $\kappa=0.02$, $\gamma=40$, $\alpha=0.05$, $1/\delta_0^2=5000$ ($\delta_0=0.014$) and $c_0=43.6/0.00035=124600$ wppm. Therefore the lines in the plot correspond to $c/c_0=0.00035$, $0.00079$, $0.0023$ and $0.0075$ }
\label{k5}
\end{figure}

In fig. (\ref{k5}) we compare the result from our model to the experimental data presented in ref. \cite{Virk71}

A comment is in order about the value of $c_0$ used to draw these plots. It is clear that the value of $c_0$ we are using corresponds to a very high concentration, probably higher than any used in actual experiments \cite{Ber78}. However, our $c_0$ does not represent the onset concentration. As shown in fig. (\ref{k5}) and will be seen again in the following figures, substantial drag reduction is seen at a concentration of $10^{-3}c_0$, and so a high value for $c_0$ is to be expected. This said, it is clear that the multiplicity of parameters and the lack of independent derivation and/or determination of at least some of them is a weakness of our model and an area for further work.

To analyze the dependence of the friction factor on concentration, it is convenient to introduce the fractional drag reduction $R_F$

\be
R_F=1-\frac f{f_0}
\label{rf}
\te
$R_F\to 0$ as $c\to 0$ by definition. When $c\to\infty$, on the other hand, it reaches a finite asymptotic value $R_{F,max}$.

We also introduce the intrinsic drag reduction $R_I$

\be
R_I=\frac {R_F}c
\label{ri}
\te
$R_I$ goes to zero when $c\to \infty$, but when $c\to 0$ it reaches a finite value $R_{I,0}$. Following Virk \cite{Virk67}, we define $G=R_{I,0}c/R_{F,max}$ and $D=R_I/R_{I,0}$. Then the following empirical relation holds

\be
D=\frac1{1+G}
\label{gamadelta}
\te

\begin{figure}[htp]
\centering
\includegraphics[scale=.8]{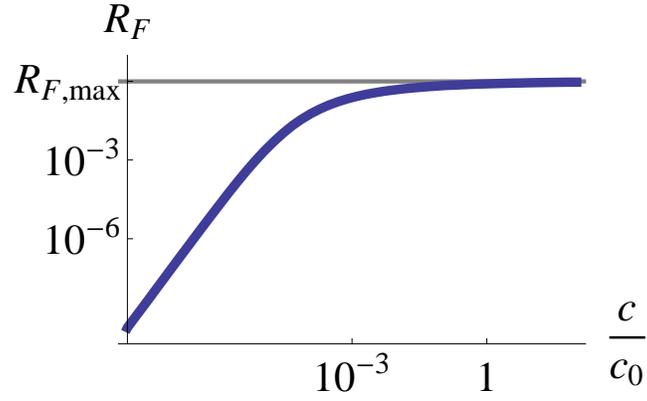}
\caption[k6] {[Color online] The fractional drag reduction for $\mathrm{Re}=10^5$.}
\label{k6}
\end{figure}

\begin{figure}[htp]
\centering
\includegraphics[scale=.8]{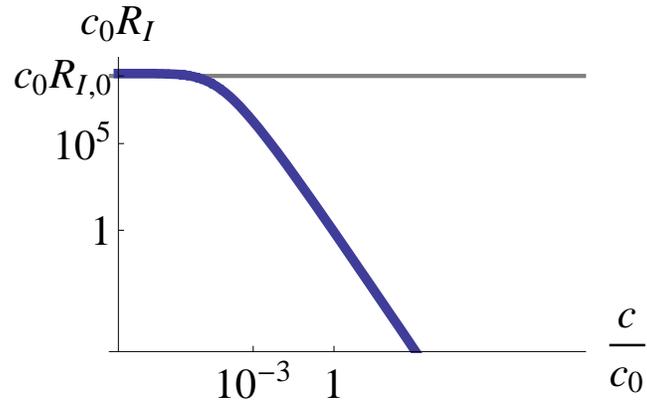}
\caption[g9] {[Color online] The intrinsic drag reduction (multiplied by $c_0$) for $\mathrm{Re}=10^5$. }
\label{k7}
\end{figure}

\begin{figure}[htp]
\centering
\includegraphics[scale=.8]{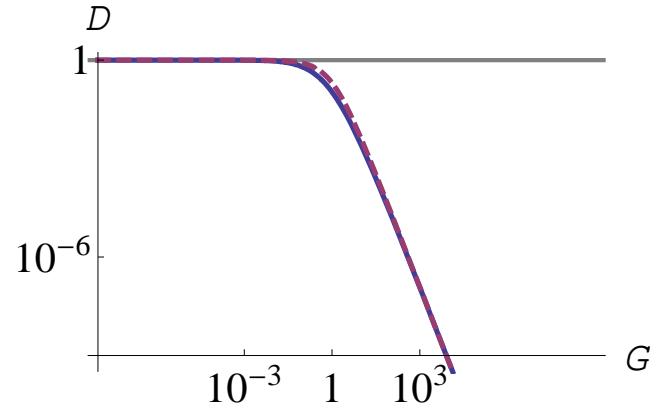}
\caption[k8] {[Color online] $D$ vs $G$ for $\mathrm{Re}=10^5$. The full curve is derived from our model, the dashed curve is the empirical relationship eq. (\ref{gamadelta}).}
\label{k8}
\end{figure}

We plot these quantities in figs. (\ref{k6}), (\ref{k7}) and (\ref{k8}). The agreement of this last figure to eq. (\ref{gamadelta}) is remarkable. This shows that our model not only predicts there will be a maximum drag asymptote, but also reproduces the full concentration dependence.

\section{Final remarks}

In this paper we have shown that, given a suitable expression for the turbulent spectrum in the absence of the polymer, there is a deformation of it that reproduces both the maximum drag reduction asymptote and the concentration dependence of the intrinsic drag reduction.

Our treatment is admittedly not a self-contained derivation of the friction factor; for once, the model allows for a large number of  parameters which are not determined independently, but simply chosen to fit experimental data regarding the friction factor itself. Granted this, we believe each step in our argument is well motivated.  The discussion of the polymer, for example, is based in what essentially is a Hartree approximation to the FENE-P model, and as such stands on a well trodden (theoretical) path.

One of the most remarkable features of the drag reduction phenomenon is the universality of the maximum drag reduction asymptote. In our model, universality obtains from the fact that we assume that, when Reynolds numbers are high, the polymer relaxation $\tau$ is determined by the eddy revolving time alone, independently of polymer characteristics (which do play a role at lower Reynolds number). This assumption is inspired in Lumley's ``time'' criterion \cite{Lum69}. However the Lumley criterion refers to the characteristic time $\rho\nu /\tau_0$, and also the way the polymer characteristic time is defined is different from ours.

The quantitative fit to the Virk asymptote depends on the other hand on the assumption that the lifetime of a velocity fluctuation grows with Reynolds number as in eq. (\ref{sigmainertial}), with the dimensionless factor eq. (\ref{nucero}). This condition follows from the requirement that both the physical and the equivalent (linear) flows share the same Reynolds number at large scales. Once it is accepted, it follows that the random driving must also weaken with Reynolds number, as could be expected from fluctuation-dissipation considerations \cite{McCKiy05}.

Overall, we believe the results of this paper are a success for the MTM, complementing earlier studies of the friction factor in two-dimensional turbulence \cite{GutGol08}. We offer them as a simple theoretical template for more fundamental approaches.

\section*{Acknowledgments}

This work is supported by University of Buenos Aires (UBACYT X032), CONICET and ANPCyT.

\section*{Appendix: The FENE-P model in the outer region}
In this appendix we shall discuss the arguments behind the contention that the polymer does not affect the flow in the outer region. We thus neglect the fluctuating velocity and assume $\mathbf{U}^p=U\left(r\right)\hat{z}^p$, $P=-p_0 z$, and $\Lambda^{pq}=\Lambda^{pq}\left(r\right)$. The left hand side of eq. (\ref{FENEP}) vanishes and we get six algebraic equations

\be
t_{S}  \left(T^{zz}-T^{zz}_{eq}\right)=2U'\Lambda^{zr}
\label{zz}
\te
where $U'=dU/dr$

\be
t_{S} T^{rz}=U'\Lambda^{rr}
\label{rz}
\te

\be
t_{S} T^{\theta z}=U'\Lambda^{\theta r}
\label{tetaz}
\te

\be
T^{rr}-T^{rr}_{eq}=T^{\theta\theta}-T^{\theta\theta}_{eq}=T^{\theta r}=0
\label{rr}
\te
Therefore

\be
\Lambda^{\theta r}=\Lambda^{\theta z}=0
\label{tetar}
\te

writing

\be
T^{pq}_{eq}=\omega_{free}^2\Lambda_{eq}g^{pq}
\label{teq}
\te
where $g^{pq}=\:diag\:\left(1,1,1/r^2\right)$ we get

\be
\Lambda^{rr}=\frac{\Lambda_{eq}}{h\left(\Lambda\right)}
\label{lambdarr}
\te

\be
\Lambda^{\theta\theta}=\frac{\Lambda_{eq}}{r^2h\left(\Lambda\right)}
\label{lambdatt}
\te

\be
\Lambda^{rz}=\frac{U'\Lambda_{eq}}{t_S\omega_{free}^2h\left(\Lambda\right)^2}
\label{lambdarz}
\te

\be
\Lambda^{zz}=\frac{\Lambda_{eq}}{h\left(\Lambda\right)}+\frac{2U'^2\Lambda_{eq}}{t_S^2\omega_{free}^4h\left(\Lambda\right)^3}
\label{lambdazz}
\te
Plus the consistency condition

\be
h\left(\Lambda\right)\frac{\Lambda}{\Lambda_{eq}}-1=\frac{2U'^2}{3t_S^2\omega_{free}^4h\left(\Lambda\right)^2}
\label{consist}
\te
In the limit when $\Lambda\approx\Lambda_{max}\gg\Lambda_{eq}$ eq. (\ref{consist}) reduces to

\be
h\left(\Lambda\right)^3=\frac{2U'^2\Lambda_{eq}}{3t_S^2\omega_{free}^4\Lambda_{max}}
\label{consist2}
\te
Given this form of the deformation tensor, the only nontrivial Navier-Stokes equations is the $z$ equation, which yields

\be
\frac{d}{d\xi}\frac1{\xi}\frac{d}{d\xi}\xi\frac{d\tilde{U}}{d\xi}=0
\label{500}
\te
where

\be
\frac{d\tilde{U}}{d\xi}=\frac{dU}{d\xi}+\frac{cR}{\nu}T^{zr}
\label{501}
\te
We see that this is the same equation as without the polymer, only the fluid velocity gradient is ``corrected'' by a term

\be
\frac{cR\omega_{free}^2h\left(\Lambda\right)}{\nu}\Lambda^{zr}=\frac{c}{\nu}\frac{\Lambda_{eq}}{t_Sh\left(\Lambda\right)}\frac{dU}{d\xi}=\frac{c}{\alpha c_0}\frac{\Lambda_{eq}}{\Lambda_{max}h\left(\Lambda\right)}\frac{dU}{d\xi}
\label{502}
\te
Even if $1/\alpha =20$, in the relevant regime all $c/c_0$, $\Lambda_{eq}/\Lambda_{max}$ and $1/h$ are very small. So the correction to $dU/d\xi$ may be disregarded for all practical purposes.

\end{document}